\documentclass[a4paper,preprint,prb]{revtex4}

\usepackage{graphicx}
\usepackage{dcolumn}
\usepackage{bm}
\usepackage{color}

\newcommand{\eqa}{\begin{equation}}
\newcommand{\eqz}{\end{equation}}
\newcommand{\eqma}{\begin{eqnarray}}
\newcommand{\eqmz}{\end{eqnarray}}

\begin{document}
\newcommand{\e}{{\em e}~}
\title{Performance of ab initio and density functional methods for conformational equilibria of C$_n$H$_{2n+2}$ alkane isomers (n=4--8)}
\author{David Gruzman, Amir Karton, and Jan M. L. Martin*}
\affiliation{Department of Organic Chemistry, 
Weizmann Institute of Science, 
IL-76100 Re\d{h}ovot, Israel}
\email{gershom@weizmann.ac.il}

\date{Draft revision \today. Received April 21, 2009. MSID: {\bf jp-2009-03640h} [Walter Thiel issue] }

\begin{abstract}
Conformational energies of n-butane, n-pentane, and n-hexane have been calculated at the CCSD(T) level and at or near the basis set limit. Post-CCSD(T) contribution were considered and found to be unimportant. The data thus obtained were used to assess the performance of a variety of density functional methods. Double-hybrid  functionals like B2GP-PLYP and B2K-PLYP, especially with a small Grimme-type empirical dispersion correction, are capable of rendering conformational energies of CCSD(T) quality. These were then used as a `secondary standard' for a larger sample of alkanes, including isopentane and the branched hexanes as well as key isomers of heptane and octane. Popular DFT functionals like B3LYP, B3PW91, BLYP, PBE, and PBE0 tend to overestimate conformer energies without dispersion correction, while the M06 family severely underestimates GG interaction energies. Grimme-type dispersion corrections for these overcorrect and lead to qualitatively wrong conformer orderings. All of these functionals also exhibit deficiencies in the conformer geometries, particularly the backbone torsion angles.
The PW6B95 and, to a lesser extent, BMK functionals are relatively free of these deficiencies. 

Performance of these methods is further investigated to derive conformer ensemble corrections to the enthalpy function, $H_{298}-H_0$, and the Gibbs energy function, ${\rm gef}(T)\equiv - [G(T)-H_0]/T$, of these alkanes. These are essential for accurate computed heats of formation of especially the larger species, as the corrections for these are several times the expected uncertainty in modern computational thermochemistry methods such as W4 theory. While $H_{298}-H_0$ is only moderately sensitive to the level of theory, ${\rm gef}(T)$ exhibits more pronounced sensitivity. Once again, double hybrids acquit themselves very well.

The effects of zero-point energy and of nonfactorizable rovibrational partition functions have been considered.
 \end{abstract}

\maketitle

\section{Introduction}

The existence of multiple conformers for n-butane and higher n-alkanes has been known since the pioneering work of Pitzer.\cite{Pitzer1940} (See the introductions to Refs.\cite{Smi96,Allinger97,Deleuze02} for a detailed bibliography.) The importance of alkanes in particular --- as basic building blocks of organic chemistry and as constituents of fossil fuels --- requires no further elaboration, nor indeed does that of aliphatic chains in general --- as moieties of lipids, of polymers like polyethylene, or of nanosystems.

Modern high-accuracy theoretical thermochemistry methods, such as W4 theory developed at Weizmann\cite{W4,W4.4} and HEAT developed by a multinational consortium of researchers\cite{heat1,heat2,heat3} are capable of calculating bottom-of-the-well total atomization energies (TAE$_e$ values) with 95\% confidence intervals of 1 kJ/mol (0.24 kcal/mol) or less. For alkanes and other systems devoid of low-lying excited electronic states, the chief factors limiting accuracy of calculated total atomization energies (TAE$_0$) or heats of formation ($\Delta H^\circ_{f,0}$) at absolute zero are actually not of an electronic structure nature so much as the zero-point vibrational energies (ZPVEs), particularly the anharmonic corrections in them. At finite temperatures, this is compounded by the necessity of reliable heat content functions or enthalpy functions ($H_{298}-H_0$). By way of illustration, a component breakdown of heat content functions for a number of lower alkanes is presented in Table \ref{tab:TRC}.  For n-octane, the conformer contribution is seen to reach 1.05 kcal/mol. Clearly, when working in the kJ/mol accuracy region, one neglects such 
contributions at one's peril.

Smith and Jaffe\cite{Smi96} studied conformational energies 
of n-butane and the central torsion of n-hexane in considerable detail. For n-butane, they 
obtained a trans-gauche energy of 0.59 kcal/mol at the CCSD(T)/cc-pVTZ//MP2/6-311G(2df,p) level. After correction, this was within 0.05 kcal/mol of the 
then-latest experimental value by Herrebout et al.\cite{Durig}

Allinger et al.\cite{Allinger97} carried out a `focal point'\cite{focalpoint1,focalpoint2} convergence
study on the butane conformer energies and torsional barriers. Their best
estimate for the trans-gauche energy was 0.62 kcal/mol, just 0.04 kcal/mol lower than the very recent measurement by Balabin,\cite{Balabin2} 0.660 $\pm$ 0.022 kcal/mol.

A similar `focal point' study for n-pentane
was published by Salam and Deleuze\cite{Deleuze02} in 2002. Their best estimates
for the relative energies of the TG (trans--gauche), GG (gauche$^+$--gauche$^+$), and GX$^-$ (gauche$^+$--perpendicular$^-$) conformers are
0.621, 1.065, and 2.917 kcal/mol, respectively. Very recently, the TG and GG conformer energies were measured by Balabin\cite{Balabin2} as
0.618$\pm$0.006 and 0.940$\pm$0.020 kcal/mol, respectively. Note that the GG conformer is about 0.3 kcal/mol more stable than what one would expect from naively adding two TG energies: the GG conformer benefits from a mild dispersive stabilization (see, e.g., \cite{Osawa91,Balabin2}).

\begin{figure}
\includegraphics[width=15cm,angle=0]{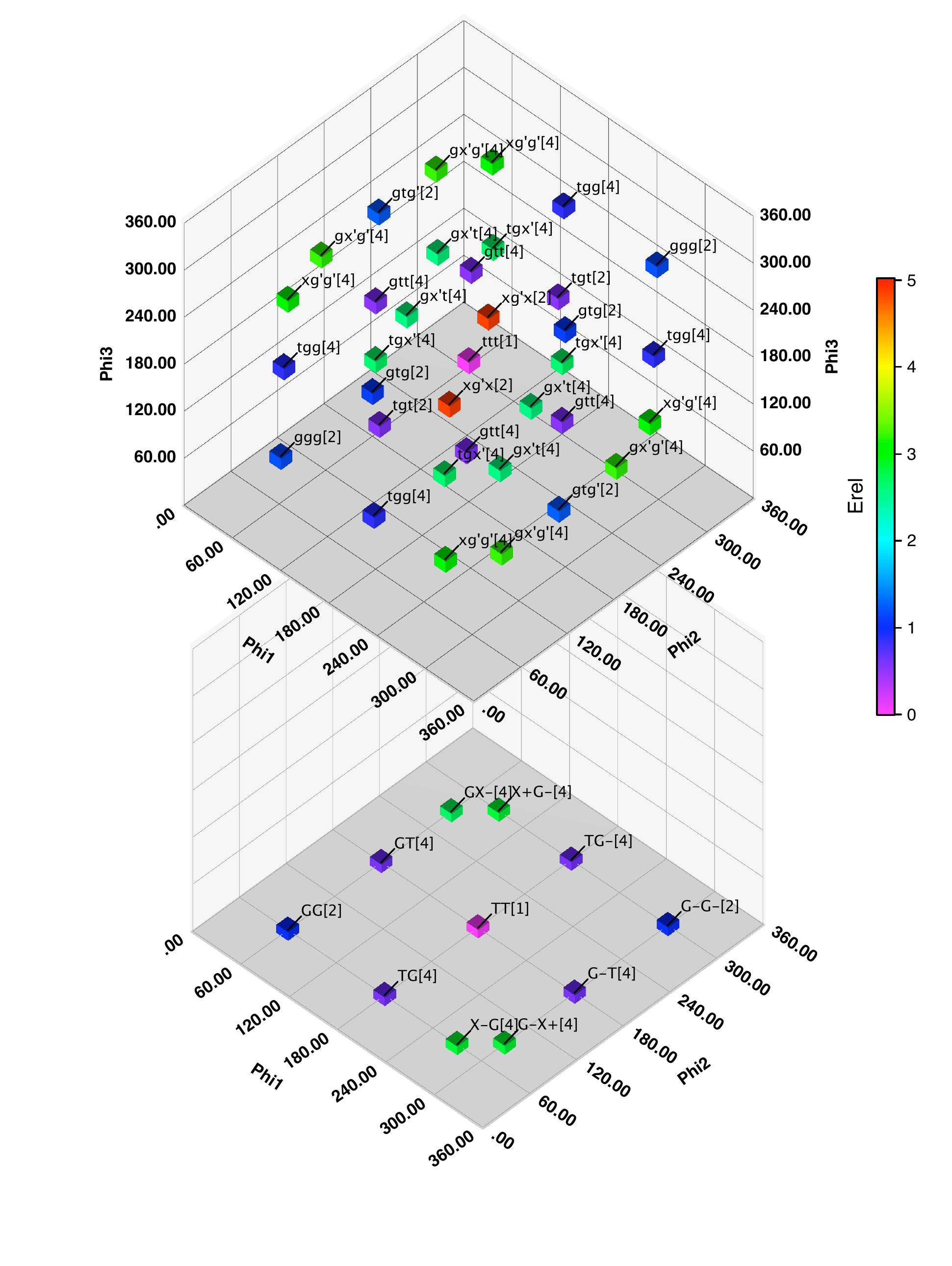} \\
\caption{Graphical representation of the conformers of $n$-pentane (bottom) and $n$-hexane (top). The more 
purple the marker, the lower the conformer is in energy. Degeneracies are noted in square parentheses.\label{fig:figure1}}
\end{figure}

Tasi et al.,\cite{Tasi1} in their paper on the enumeration of conformers in
n-alkanes, discuss n-pentane at some length. Figure 1 in their paper is an 
energy landscape of n-pentane in terms of the two CCCC torsion angles. 
(In the remainder of this paper, we will adopt their notation for conformers:
g$^\pm$ for gauche torsion angles around $\pm$60$\deg$, x$\pm$ for `cross' or
`perpendicular' angles around $\pm$90$\deg$, and t for trans.)
It is seen there (as well as in the lower part of Figure \ref{fig:figure1} in the present paper) that the landscape has eleven minima: the global tt minimum,
two equivalent gg minima (g$^+$g$^+$ and g$^-$g$^-$), four equivalent tg minima (t$^+$g$^+$, 
t$^+$g$^-$, g$^+$t, and g$^-$t), and finally four equivalent gx$^-$ minima (g$^+$x$^-$, g$^-$x$^+$,
x$^+$g$^-$, and g$^-$g$^+$). The gx- conformer is often, confusingly, labeled g$^+$g$^-$ in older literature:
in fact, the actual g$^+$g$^-$ structure is a transition state for a shallow barrier
between equivalent g$^+$x$^-$ and x$^+$g$^-$ structures (and similarly for g$^-$g$^+$ between
x$^-$g$^+$ and g$^-$x$^+$).

Because of this latter phenomenon (first pointed out, to the best of our knowledge, by
Osawa and coworkers,\cite{Osawa91}) higher $n$-alkanes have more conformers than
would be expected by naive "$3^n$" enumeration based on trans/gauche$^+$/gauche$^-$
products. This latter approach does lead to the correct four conformers for
$n$-pentane (even as it mislabels the energetically highest one), but only
yields ten unique conformers on the n-hexane surface while in reality 
there are twelve. A graphical representation of the conformer space of n-hexane
can be seen in the upper part of Figure \ref{fig:figure1}.

While the ttt, gtt, tgt, tgg, gtg, gtg$^-$, and ggg conformers also occur
in the naive $3^n$ enumeration and the naive gg$^-$g conformer could be regarded
as a `rounded-off' equivalent of the actual xg$^-$x conformer, the naive ggg$^-$ and
gg$^-$t conformers actually each correspond to nonequivalent pairs, gx$^-$t/xg$^-$t
and gx$^-$g/ggx$^-$, respectively.

Tasi et al.\cite{Tasi1,Tasi2} defined rules for a more correct conformer
enumeration, based on pruning a $5^n$ search down by excluding `forbidden'
sequences that exhibit excessive sterical hindrance. The rules can be 
summarized as follows:
\begin{itemize}
\item g$^+$g$^-$, g$^-$g$^+$,  x$^+$x$^-$, and x$^-$x$^+$ are forbidden. Instead, g$^+$x$^-$/x$^+$g$^-$ and g$^-$x$^+$/x$^-$g$^+$ pairs occur.
\item gx$^-$g, xgx$^-$, and their isomorphs are forbidden
\item xg$^-$g$^-$x and its isomorph x$^-$ggx$^-$ are forbidden
\end{itemize}

Under these rules, 30 and 95 nonequivalent conformers occur for n-heptane and
n-octane, respectively.

In the present work, we will first obtain benchmark ab initio conformer energies for $n$-butane, $n$-pentane, and $n$-hexane, using large basis set CCSD(T) as a `primary standard'. We will then show that double-hybrid functionals\cite{B2-PLYP,B2GP-PLYP} supplemented by an empirical dispersion correction yields results of essentially the same quality, and will proceed to use these as a `secondary standard'. Next, we will consider the performance of a variety of density functional methods for the conformers of the pentanes, hexanes, and selected heptanes and octanes. Finally, we will address the quality of computed thermodynamic corrections both by the more rigorous and the more approximate methods.

\section{Computational methods}

\subsection{Electronic structure}

All calculations were carried out using MOLPRO 2008.1\cite{molpro} or a locally modified version of Gaussian 03 rev. E.01\cite{g03} running on the Martin group Linux cluster at Weizmann. Gaussian was used for all DFT calculations as well as for MP2 optimizations: MOLPRO was used for the CCSD(T) calculations. Some post-CCSD(T) calculations were carried out using MRCC.\cite{MRCC}

The following DFT functionals were considered (grouped by rungs on the `Jacob's Ladder' of Perdew:\cite{JacobsLadder})
\begin{itemize}
\item second-rung (i.e., GGAs): BLYP,\cite{B88ex,LYPc} PBE\cite{PBE}
\item third-rung (i.e., meta-GGAs): M06L\cite{M06-L}
\item imperfect fourth rung (i.e., hybrid GGAs): B3LYP,\cite{B3,B3LYP} B3PW91,\cite{B3,PW91c} PBE0\cite{PBE0}
\item full fourth rung (i.e., hybrid meta-GGAs): B1B95,\cite{B88ex,B95c} PW6B95,\cite{PW6B95} BMK,\cite{BMK} M06,\cite{M06} M06-2X\cite{M06}
\item fifth rung: the double hybrids B2-PLYP,\cite{B2-PLYP} B2GP-PLYP\cite{B2GP-PLYP} and B2K-PLYP\cite{B2K-PLYP}
\end{itemize}

Most wavefunction ab initio calculations were carried out using the cc-pVnZ\cite{Dun89} and aug-cc-pVnZ\cite{Ken92} basis sets of Dunning and coworkers. In the course of this paper, we will refer to the cc-pVnZ basis set by the PVnZ acronym, and to the combination of aug-cc-pVnZ on carbon with regular cc-pVnZ on hydrogen by the AVnZ acronym (n=D,T,Q).

Most DFT calculations were carried out using the Jensen pc-2 basis set.\cite{pc-2}

Dispersion corrections for the DFT energies (denoted by the suffix "-D") were applied using our implementation of Grimme's expression:\cite{Grimme1,Grimme2} 
\begin{equation}
E_{disp} = -s_{6} \sum_{i=1}^{N_{at}-1} \sum_{j=i+1}^{N_{at}} \frac{C_{6}^{ij}}{R_{ij}^{6}}f_{\rm dmp} \left(R_{ij} \right)\label{eq:grimme}
\end{equation}
where the damping function is taken as
\begin{equation}
 f_{\rm dmp} \left( R_{ij} \right) = \left[ 1 + \exp \left( -\alpha ( \frac{R_{ij}}{s_RR_{r}} -1 )\right)\right]^{-1}\label{eq:damping}
\end{equation}
and $C_{6}^{ij}\approx\sqrt{C_{6}^{i}C_{6}^{j}}$, $R_{r}=R_{{\rm vdW},i}+R_{{\rm vdW},j}$  is the sum of the van der Waals radii of the two atoms in question, and the specific numerical values for the atomic Lennard-Jones constants $C_{6}^{i}$ and the van der Waals radii (in this case, 1.452 \AA\ for C and 1.001 \AA\ for H) have been taken from Ref.\cite{Grimme1}. The length scaling $s_R$=1.0 and hysteresis exponent $\alpha$=20.0 were set as in Ref.\cite{Grimme2}.

This expression is left with a single functional-dependent empirical parameter, namely the prefactor $s_6$. This was taken from Refs.\cite{Grimme1,Grimme2} for BLYP, B3LYP, and PBE, from Ref.\cite{B2GP-PLYP} for the double hybrids, and from Ref.\cite{alkanes} for the remaining functionals. These were, for the most part, optimized against the S22 benchmark set of weakly interacting systems.\cite{Hobza06}

No corrections for intramolecular BSSE were made: instead, we elected to use basis sets sufficiently large ($spdf$ or $spdfg$ quality) that this should not be an issue on the accuracy scale of interest to us presently. We note that Balabin\cite{Balabin1} studied intramolecular BSSE for trans/gauche butane and selected hexanes in considerable detail.

\subsection{Other aspects}

The initial conformer structures were generated by stepping the CCCC dihedral
angles in 30 degree increments, running molecular mechanics optimizations on 
all structures generated, and collating equivalent structures. In this manner,
both the unique conformer structures and their degeneracies are obtained in 
an automated fashion.

The statistical thermodynamic corrections were then obtained by state summation
according to the method of Pitzer and Brewer.\cite{PitBrew} That is, the 
partition function and its first two moments are obtained as follows:
\begin{eqnarray}
Q &=& \sum_i{d_i\exp\left(-\frac{E_i-E_0}{RT}\right)}\\
Q' &=& \sum_i{d_i\left(\frac{E_i-E_0}{RT}\right)\exp\left(-\frac{E_i-E_0}{RT}\right)}\\
Q'' &=& \sum_i{d_i\left(\frac{E_i-E_0}{RT}\right)^2\exp\left(-\frac{E_i-E_0}{RT}\right)}
\end{eqnarray}
where $E_i$ and $E_0$ are the energies of state $i$ and the ground state,
respectively (in molar units), R is the gas constant (replace by the Boltzmann
constant $k$ if molecular units are preferred for the energies), $d_i$ is the
degeneracy of the state, and T is the temperature in Kelvin.
The various thermodynamic functions are then obtained as:
\begin{eqnarray}
{\rm gef}(T)\equiv -\frac{G_T-H_0}{T}&=& R \ln Q\\
{\rm hcf}(T)=H_T-H_0&=& RT Q'/Q\\
S(T)=R\left(\frac{Q'}{Q}+\ln Q\right)\\
C_p(T)=R\left(\frac{Q''}{Q}-\left(\frac{Q'}{Q}\right)^2\right)
\end{eqnarray}

A slight complication is introduced when the internal partition function is
not assumed to be factorizable, i.e., each of the conformers is allowed to
have distinct rotational, vibrational, and (ignored in this work) electronic
partition functions. Then the following product rules apply (Eqs. 1-3 in 
Ref.\cite{nh2}):
\begin{eqnarray}
Q=Q_0+\sum_i{\exp(-x_i)Q_i}\\
Q'=Q'_0+\exp(-x_i)\left[x_iQ_i+Q'_i \right]\\
Q''=Q''_0+\exp(-x_i)\left[x_i^2Q_i+Q''_i + 2 x_iQ'_i\right]
\end{eqnarray}
where the shorthand $x_i\equiv(E_i-E_0)/RT$ was applied.

While the expression for $C_p(T)$ is too clumsy for practical use, the 
following equations express ${\rm gef}(T)$ and ${\rm hcf}(T)\equiv H_T-H_0$ in terms of the
quantities for the individual conformers:
\begin{eqnarray}
{\rm gef}(T)={\rm gef}(T)_0 + R \ln\left[1+\sum_i{d_i\exp(-x_i)\frac{Q_i}{Q_0}}\right]\\
{\rm hcf}(T)=\frac{{\rm hcf}_0(T)+\sum_i{\exp(-x_i)\frac{Q_i}{Q_0}[RTx_i+{\rm hcf}_i(T)] } }{1+\sum_i{\exp(-x_i)\frac{Q_i}{Q_0}}}
\end{eqnarray}

\section{Results and discussion}

\subsection{Wavefunction ab initio}

The wavefunction ab initio results are gathered in Table \ref{tab:ab-initio}. (Results in this section apply to MP2/cc-pVTZ reference geometries.) 

\subsubsection{n-butane and n-pentane}

Applying W1 extrapolation\cite{W1} to CCSD(T)/PV\{D,T\}Z and CCSD/PV\{T,Q\}Z energies results in 0.598 kcal/mol. This is basically identical to our best result, 0.596 kcal/mol, which is obtained by W2 extrapolation\cite{W1} to CCSD(T)/PV\{T,Q\}Z and CCSD/PV\{Q,5\}Z data.
At the CCSD(T)/A'VQZ level, the trans-gauche conformer separation is 0.609 kcal/mol. The CCSD(T)/PVQZ result is insignificantly different (0.606 kcal/mol), as is the CCSD(T)/A'VTZ result (0.613 kcal/mol). CCSD(T)/PVTZ still comes quite close, at 0.588 kcal/mol (less than 0.01 kcal/mol below our best number): smaller basis sets exhibit more significant deviations (0.721 kcal/mol for A'VDZ, 0.693 kcal/mol for PVDZ). 
Our best value is in excellent agreement with the experimental value of 0.67$\pm$0.10 kcal/mol of Herrebout et al.\cite{Durig} as well as the best calculation of 0.62 kcal/mol by Allinger et al.\cite{Allinger97}, and the 0.628 kcal/mol obtained by Klauda et al.\cite{Klauda} at what they term the MP2:CC level. (This is their shorthand for a procedure that is essentially MP2/PVQZ + [CCSD(T)/PVDZ -- MP2/PVDZ] at MP2/PVDZ reference geometries.)

Let us now considering the n-particle convergence with the cc-pVTZ basis set. SCF, as expected, overestimates the separation at 1.138 kcal/mol, while MP2 slightly overcorrects at 0.561 kcal/mol. However, the accidental very good agreement with the CCSD(T) number of 0.588 kcal/mol results from a fortunate error compensation: the gap widens to 0.664 kcal/mol at the MP3 level, to 0.635 kcal/mol at the MP4(SDQ) level, and to 0.653 kcal/mol at the CCSD level.

We considered the effect of post-CCSD(T) correlation by carrying out CCSDT(Q)/cc-pVDZ(no $p$ on H) calculations for both the trans and the gauche structures. 
Connected quadruple excitations (Q) and higher-order triple excitation effects (i.e., the CCSDT -- CCSD(T) difference) are found to stabilize the gauche conformer by just 0.002 kcal/mol each: we conclude that post-CCSD(T) effects are insignificant on the accuracy scale we are interested in.

In the case of n-pentane, our best (W1h-val-type) estimates for the conformer energies of \{TG,GG,and GX$^-$\} relative to TT are 
\{0.614,0.961,2.813\} kcal/mol, only marginally different from numbers are obtained from (T) contributions with the AVTZ basis set and CCSD contributions with the PVQZ basis set, \{0.605,0.954,2.810\} kcal/mol.
As expected, the CCSD(T)/AVTZ numbers are close at
\{0.586,0.946,2.817\} kcal/mol, as are the CCSD(T)/PVTZ numbers at \{0.581,0.919,2.767\} kcal/mol, even though the latter hint at an undershooting problem that will become clearer for n-hexane.
 The same general trends as for n-butane apply to n-particle convergence: MP2 slightly overcorrects, CCSD spaces conformers too wide. The MP2 error is largest for the GG conformer; the (T) contributions for the GG and GX$^-$ conformer energies are noticeably larger than for TG. The TG and GG values are in excellent agreement with the latest measurements by Balabin,\cite{Balabin2} 
0.618$\pm$0.006 and 0.940$\pm$0.020 kcal/mol, respectively.

SCS-MP2\cite{SCS-MP2} yields conformer relative energies that are basically indistinguishable from CCSD.

Klauda et al.\cite{Klauda} report MP2:CC numbers of \{0.622,0.985,2.846\} kcal/mol, which are quite close to our higher-level data.
Our own MP2:CC calculation at our higher-level (MP2/cc-pVTZ) geometries yields slightly different numbers.

Our best estimates are somewhat different from those of Salam and Deleuze: \{0.621,1.065,2.917\} kcal/mol. Detailed analysis suggests that these differences are primarily due to their use of B3LYP/6-311++G(d,p) reference geometries. We shall see below that the B3LYP functional is inadequate for this purpose, not just in relative energies which are \{0.855,1.604,3.360\} at the B3LYP/6-311++G(d,p) level but also in terms of the calculated geometries: as can be seen in Table \ref{tab:dihedral}, the backbone torsion angles for the GG and GX$^-$ conformers are off by 5--6 degrees.

SCS-MP2 tracks the CCSD numbers quite closely, while SCS-CCSD clearly overcorrects for the (T) contribution. 

The TT--GG energy difference appears to be the most sensitive to the level of theory, followed by the TT--GX$^-$ difference.

\subsubsection{n-hexane}

n-hexane, with its twelve conformers, becomes a bit unwieldy to compare in terms of individual conformer energies. One could instead focus on the calculated conformer contribution to the enthalpy function, which is the quantity that interests us most from an utilitarian point of view. However, by construction, this will be most affected by the low-lying gtt and tgt conformers. 

Let us first consider the relative energies of the conformers at the W1h-val level. Obviously, the ttt conformer is lowest in energy, followed by nearly isoenergetic gtt and tgt conformers. Somewhat higher up is tgg, which is markedly more stable than gtg. gtg$^-$ is about 0.13 kcal/mol less stable than its cousin gtg, while ggg finds itself sandwiched between them. The remaining five conformers all have at least one `pentane interference' interaction: gxt and tgx- are nearly isoenergetic (and about 1.4 kcal/mol less stable than ggg), as is the less stable xg$^-$g$^-$ and gxg$^-$ pair, while the least stable conformer is xg$^-$x. Visual inspection reveals (see Electronic Supporting Information) that this latter conformer is basically a `helix', one end nearly coiling back over the other.

Klauda et al.\cite{Klauda} calculated MP2:CC relative energies for selected conformers: tgt 0.600, tgg 0.930, tgg$^-$ 2.740, gtg 1.180, gtg$^-$ 1.320 kcal/mol. We computed a complete set (Table II) at our own reference geometries, and find MP2:CC to agree with our W1h-val data to within about 0.01 kcal/mol. In contrast, CCSD(T)/cc-pVTZ data are biased downward by as much as 0.1 kcal/mol, presumably due to basis set superposition error.

\subsection{Density functional methods}

In order to basically eliminate the basis set as a factor in our comparison, we have used the extended pc-2 basis set throughout. Except for the double-hybrid results, a full optimization was carried out for every functional.

Let us begin by considering the pentane and hexane conformers. The energetic data are collected in Table \ref{tab:dft-ener}.

A few interesting features emerge. First, the percentage of Hartree-Fock exchange does not seem to be all that important: the BLYP/B3LYP pair on the one hand, and the PBE/PBE0 pair on the other hand, yield basically the same answers. 

Second, while conventional DFT functionals (such as B3LYP, PBE0, B3PW91,...) tend to overestimate conformer gaps (progressively moreso as one goes higher up the conformer ladder), the M06 family of Truhlar underestimates them. Of particular note is the situation in pentane, where M06 wrongly finds the TG and GG conformers to be energetically nearly degenerate. Similarly distorted energetic pictures are seen for hexane conformers: we note, from considering the Grimme dispersion corrections for the various structures, that the GG interaction is favored by dispersion, and it appears that the M06 family overestimates that impact. The PW6B95 and BMK functionals, on the other hand, surprisingly do a much better  job.

Third, it was previously noted\cite{Chirality} that Grimme-type empirical dispersion corrections considerably improved DFT relative energies for selected conformers of 4-ethyl-4-methyloctane. 
Comparing with a larger sample of higher-level reference data, however, we find that Grimme-type corrections appear to overcorrect for the conventional DFT functionals. Ad hoc reoptimization of $s_6$ prefactors revealed that, for alkane conformer energies, one would like a prefactor about 0.4-0.5 smaller than the generic optimum. For the M06 family, {\em ad hoc} optimized $s_6$ values are actually negative, which can be seen as ``undoing'' an overcorrection for dispersion.

Detailed inspection of dispersion correction contributions for the simplest case (the trans-gauche equilibrium in butane) reveals that dispersion interactions between the terminal CH$_3$ group and the CH$_2$ group in $\beta$ position relative to it ("1,3" interactions, if one likes) actually favor the trans conformer (as it has two $\beta$ hydrogens in close proximity rather than one), but that the gauche conformer enjoys much more favorable "1,4" dispersion interactions. The 1,3-interactions are in the distance range where the damping function, Eq.(\ref{eq:damping}), rapidly turns over, making the overall correction quite sensitive to its details. In addition some double-counting with the DFT correlation functional is inevitable.

Fourth, the double-hybrids B2GP-PLYP and B2K-PLYP perform fairly well even without dispersion corrections, and very well when supplemented with their standard dispersion corrections ($s_6$=0.40 for B2GP-PLYP-D and 0.30 for B2K-PLYP-D). Ad hoc optimization results in $s_6$ values that, unlike for the conventional functionals, are only slightly smaller than the standard values: $s_6$=0.28 for B2GP-PLYP and $s_6$=0.22 for B2K-PLYP. (Note that, if CCSD(T)/cc-pVTZ reference data were used instead for calibration, higher $s_6$ values of 0.32 for B2GP-PLYP and 0.26 for B2K-PLYP would result, which would lead to a downward bias for all conformer energies.) These results once again underline the robustness and versatility of the B2GP-PLYP and B2K-PLYP functionals. After {\em ad hoc} adjustment of $s_6$, the various double hybrids yield results of comparable quantity: in the remainder of the paper, we have somewhat arbitrarily restricted ourselves to B2K-PLYP as it requires the smallest adjustment to $s_6$, but we could have used B2GP-PLYP to equally good effect.

Fifth, the deficiencies of several functionals are not just reflected in the energetics, but also in the geometries. This is especially noticeable in the backbone torsion angles: Table \ref{tab:dihedral} contains the dihedral angles for the TG, GG, and GX$^-$ conformers of n-pentane by way of illustration.

As MP2/cc-pVTZ optimizations for all heptane and especially octane conformers would be computationally too unwieldy, we selected the PW6B95 functional for optimizing the reference geometries of the remaining conformers. Some exploratory calculations on pentane and hexane revealed that the 6-311G** basis set was adequately converged for our purposes, and that B2K-PLYP/pc-2 energetics at these reference geometries are very close to those obtained at MP2/cc-pVTZ geometries. The PW6B95/6-311G** level of theory was thus selected for the remaining conformer sets.

\subsection{n-heptane, n-octane, and the branched alkanes}

B2K-PLYP-D/pc-2//PW6B96/6-311G** relative conformer energies for all species considered in this paper can be found in Table \ref{tab:b2k-plyp}. We shall briefly survey the conformer sets here.
As an additional ``sanity check'' on our procedure, we have calculated the n-heptane conformer energies at the MP2:CC level as well. These results are compared with the B2K-PLYP-D(0.22)/pc-2 data in Table S1 of the Supporting Information. The two sets of values are in very close agreement with each other with an RMSD of just 0.04 kcal/mol. Refitting $s_6$ to this larger sample of 30 conformers revealed no significant change: $s_6$=0.224 RMSD, $s_6$=0.214 when fitted to RMSRelD, between which values $s_6$=0.22 is a good compromise. While n-heptane still has a pronounced ``band gap'' of sorts between the conformers involving only t and g interactions and the conformers involving x$^\pm$g$^\mp$ or g$^\pm$x$^\mp$ sequences, this gap becomes much smaller for n-octane.

Isopentane has just two conformers: the no-symmetry ground state conformer (with a "trans" backbone skeleton) and, around 0.79 kcal/mol higher, a gauche-like conformer with $C_s$ symmetry.

Isohexane (2-methylpentane) has seven conformers: these are best understood by substituting a methyl group on the four unique conformers of n-pentane. The ground-state conformer is TT; the TG and GT conformers become nonequivalent because of the methyl group; GX$^-$ and X$^-$G likewise become nonequivalent; and unlike for n-pentane, the GG conformer is actually the highest in energy here.

Isoheptane has some eighteen conformers, which are again best understood by substituting a methyl group on the 12 unique conformers of n-hexane and considering the resulting loss of  spatial degeneracy. Further details can be found in the Electronic Supporting Information.

Iso-octane, the "100\%" fixpoint on the octane scale, does not have the usual "2-methyl" backbone structure of the lower isoalkanes, but is effectively (t-butyl,isopropyl)methane. It has just three conformers, all without symmetry: in the global minimum, the iPr and tBU groups are oriented anti with respect to each other (fairly close to $C_s$ symmetry), while a "gauche" type structure is just 0.5 kcal/mol above and a third, "syn" like conformer, is found 3.3 kcal/mol above the global minimum.

3-methylpentane has six conformers.  The global minimum has $C_s$ symmetry: the other ones are best understood by considering n-pentane with a substituent in 3~position, making the fourfold degenerate TG and GX$^-$ conformer split up into nonequivalent pairs.

Biisopropyl (2,3-dimethylbutane) has just two conformers: the trans conformer with $C_{2h}$ symmetry and, less than 0.1 kcal/mol higher, the gauche conformer with $C_2$ symmetry.

Neoheptane has just three: the global minimum with $C_s$ symmetry and two asymmetric conformers at 2.3 and 2.7 kcal/mol higher. Their impact on the thermodynamic functions is minimal.

\subsection{Thermodynamic function corrections}

As pointed out in the introduction, one of the main motivations for the present study was a thermochemical one, namely the need for reliable conformer corrections to the enthalpy function and Gibbs energy function of the alkanes. Such data for the various species considered in this paper, as well as in Ref.\cite{alkanes}, can be found in Table \ref{tab:ef-gef} for various levels of theory.

As can be seen there, for the few systems where extended basis set CCSD(T) data are available, B2K-PLYP-D/pc-2 yields nearly identical results.
We thus take this level as our yardstick for the thermodynamic function corrections for the remaining species.

The enthalpy function correction --- which is what is needed for obtaining heats of formation from atomization energies, or vice versa --- exhibits fairly mild sensitivity to the level of theory. Broadly speaking, the M06 family tends to significantly underestimate the corrections while popular functionals like B3LYP tend to overestimate it. The anomalous negative sign for M06 and M06-L in the case of diisopropyl results from the wrong conformer ordering being predicted.

Sensitivity of the Gibbs energy function is rather more pronounced and behavior of the different functionals rather less systematic.

We finally address the issue of cross-coupling with zero-point and thermal corrections. Table \ref{tab:thermal-corrections}  compares B2K-PLYP-D/pc-2 thermal conformer corrections obtained in three different manners: (a) using bottom-of-the-well conformer energy differences ($\Delta E_e$); (b) using conformer energy differences at 0 K ($\Delta E_0$); (c) in addition, including individual rovibrational partition functions for all conformers (i.e., not assuming the rovibrational and conformer partition functions to be factorizable).
For want of a computationally affordable alternative, the RRHO (rigid rotor-harmonic oscillator) approximation was applied to both ZPVE and thermal corrections. (the molecular constants required were obtained at the PW6B95/6-311G** level with "ultrafine" integration grids, i.e., pruned (99,590) for energy and gradient and (50,194) for harmonic frequencies. For more on the sensitivity of harmonic frequencies to DFT integration grids, see Ref.\cite{grids}.) The thermochemical consequences of the RRHO approximation on the relative conformer energies are hard to quantify. The effects of (b) and (c) on the enthalpy function are generally quite modest for the n-alkanes (relatively speaking), but more pronounced for some of the branched alkanes, notably 3-methylpentane, isoheptane and isooctane. These general tendencies are exacerbated for the Gibbs energy function.

For n-butane through n-heptane, we considered internal rotation corrections for each individual conformer by means of the Ayala-Schlegel approximation\cite{AyalaSchlegel} . These results are given in the bottom pane of Table \ref{tab:thermal-corrections}. As can be seen there, the effect on the enthalpy functions is minimal, and that on the Gibbs energy function quite modest as well, considering that 0.1 e.u. translates to less than 0.03 kcal/mol in the free energy. We thus feel justified in not considering it for the other conformers.

Finally, one wonders about whether a CH$_2$ group equivalent could be applied to longer alkane chains. Linear regression of the corrections for n-butane through n-octane in terms of the number of backbone torsion angles reveals especially good correlation coefficients at the bottom of the well ($R^2$=0.9998 for $H_{298}-H_0$, 0.9993 for ${\rm gef}(T)$).
If zero-point energy is taken into account, we see a mild deterioration of the fit for $H_{298}-H_0$ but a somewhat more pronounced one for ${\rm gef}(T)$, while the fits including full thermal averaging become a bit noisier than desirable but still adequate for estimation purposes.

\section{Conclusions}

Conformational energies of n-butane, n-pentane, and n-hexane have been calculated at the CCSD(T) level and at or near the basis set limit. Post-CCSD(T) contribution were considered and found to be unimportant. The data thus obtained were used to assess the performance of a variety of density functional methods. Double-hybrid  functionals like B2GP-PLYP and B2K-PLYP, especially with a small Grimme-type empirical dispersion correction, are capable of rendering conformational energies of CCSD(T) quality. These were then used as a `secondary standard' for a larger sample of alkanes, including isopentane and the branched hexanes as well as key isomers of heptane and octane. Popular DFT functionals like B3LYP, B3PW91, BLYP, PBE, and PBE0 tend to overestimate conformer energies without dispersion correction, while the M06 family severely underestimates GG interaction energies. Grimme-type dispersion corrections for these overcorrect and lead to qualitatively wrong conformer orderings. All of these functionals also exhibit deficiencies in the conformer geometries, particularly the backbone torsion angles.
The PW6B95 and, to a lesser extent, BMK functionals are relatively free of these deficiencies. 

Performance of these methods is further investigated to derive conformer ensemble corrections to the enthalpy function, $H_{298}-H_0$, and the Gibbs energy function, ${\rm gef}(T)\equiv - [G(T)-H_0]/T$, of these alkanes. These are essential for accurate computed heats of formation of especially the larger species, as the corrections for these are several times the expected uncertainty in modern computational thermochemistry methods such as W4 theory. While $H_{298}-H_0$ is only moderately sensitive to the level of theory, ${\rm gef}(T)$ exhibits more pronounced sensitivity. Once again, double hybrids acquit themselves very well.

The effects of zero-point energy and of nonfactorizable rovibrational partition functions were considered, and found to be smaller than those arising from an inadequate level of theory for the conformer energies.

\section*{Acknowledgments}
Research at Weizmann was funded by the Israel Science Foundation (grant 709/05), 
the Helen and Martin Kimmel Center for Molecular Design, and the Weizmann Alternative Energy Research Initiative (AERI).
JMLM is the incumbent of the Baroness Thatcher Professorial Chair of Chemistry and a member {\em ad personam} of the Lise Meitner-Minerva Center for Computational Quantum Chemistry. 

\section*{Supporting Information}

Optimized geometries in Cartesian coordinates of all conformers discussed here, as well as visualization of same using Jmol, can be viewed at the URL 
\url{http://theochem.weizmann.ac.il/web/papers/alkane-conformers/}. PW6B95/6-311G** harmonic frequencies,  zero-point vibrational energies, and vibrational enthalpy functions of all conformers can also be found there, as can Table S-1.

\clearpage

\squeezetable
\begin{table}
\caption{Component breakdown of the theoretical enthalpy functions $H_{298}-H_0$ (kcal/mol) of several lower alkanes and comparison between theoretical and experimental values.\label{tab:TRC}}
\begin{tabular}{l|ccc|cccccc}
\hline\hline
 &  &  &  & \multicolumn{3}{c}{H$_{298}-$H$_0$}\\
 & & & & This work & \multicolumn{2}{c}{Expt.}\\
 & vibrational$^d$ & conformer$^e$ & Internal rotation$^d$ & Total &  CCCBDB$^a$\\
\hline 
ethane         & 0.41 & 0.00 & 0.05 & 2.83 & 2.84\\
propane        & 1.08 & 0.00 & 0.09 & 3.54 & 3.52\\
n-butane       & 1.86 & 0.26 & 0.13 & 4.62 &  4.61\\
n-pentane      & 2.68 & 0.47 & 0.19 & 5.7 &  5.78\\
n-hexane       & 3.51 & 0.68 & 0.24 & 6.79 &  6.86\\
n-heptane      & 4.35 & 0.89 & 0.28 & 7.87 &  7.94$^b$\\
n-octane       & 5.20 & 1.08 & 0.35 & 8.97 &  9.03$^b$\\
isobutane      & 1.81 & 0 & 0.11 & 4.29 &  4.29\\
isopentane     & 2.65 & 0.09 & 0.19 & 5.3 &  5.26\\   
neopentane     & 2.41 & 0 & 0.06 & 4.84 &  5.54$^c$\\   
isohexane      & 3.45 & 0.24 & 0.21 & 6.28 &  6.29\\
3-methylpentane& 3.50 & 0.26 & 0.24 & 6.37 &  6.23\\
diisopropyl    & 3.44 & 0.04 & 0.22 & 6.07 &  5.85\\    
neohexane      & 3.41 & 0 & 0.2 & 5.98 &  6.01\\ 
isoheptane     & 4.34 & 0.46 & 0.29 & 7.44 &  7.39$^b$\\
neoheptane     & 4.28 & 0.15 & 0.21 & 7.01 &  6.98$^b$\\
hexamethylethane&4.92 & 0 & 0.22 & 7.52 &  7.53$^b$\\
isooctane      & 5.02 & 0.16 & 0.22 & 7.77 &  7.69$^b$\\
\hline\hline
\end{tabular}
\begin{flushleft}
At room temperature, $[H_{298}-H_0]_{\rm trans+rot}$=4RT=2.37 kcal/mol for all practical intents and purposes.

$^a$ NIST CCCBDB\cite{cccbdb} unless indicated otherwise. Most of these data taken from next reference.\\
$^b$ Scott, D. W.;``Chemical Thermodynamic Properties of Hydrocarbons and related substances'', U.S. Bureau of Mines Bulletin No. 666 (US Government Printing Office, Washington, DC, 1974). Available online at \url{http://digicoll.manoa.hawaii.edu/techreports/PDF/USBM-666.pdf}. Indirectly (via TRC database) the source for most of the CCCBDB data.\\
$^c$ As shown in Ref.\cite{alkanes}, this value is erroneous.\\
$^d$ Ref.\cite{alkanes}.\\
$^e$ Present work, the values here may differ by 0.01--03 kcal/mol from the values given in the Supporting Information of Ref.\cite{alkanes} due to the use of a slightly different $s_6$ value for the B2K-PLYP-D functional (see text).\\
\end{flushleft}
\end{table}

\clearpage

\squeezetable
\begin{table}
\caption{Relative energies (in kcal/mol) of $n$-butane, $n$-pentane, and $n$-hexane conformers at MP2/cc-pVTZ reference geometries.\label{tab:ab-initio}}
\begin{tabular}{l|l|ccccc|cc|ccccc|cc}
\hline\hline
&  & \multicolumn{5}{c}{cc-pVTZ} & \multicolumn{2}{c}{aug-cc-pVTZ} & \multicolumn{5}{c}{cc-pVQZ} &  & \\
&  & HF & MP2 & SCS-MP2 & CCSD & CCSD(T) & CCSD & CCSD(T) & HF & MP2 & SCS-MP2 & CCSD & CCSD(T) & W1h-val$^a$ & MP2:CC\\
\hline
 & &\multicolumn{14}{c}{$n$-butane} \\
\hline
T & C$_{2h}$ & 0.000 & 0.000 & 0.000 & 0.000 & 0.000 & 0.000 & 0.000 & 0.000 & 0.000 & 0.000 & 0.000 & 0.000 & 0.000 & 0.000\\
G & C$_2$    & 1.138 & 0.561 & 0.657 & 0.653 & 0.588 & 0.679 & 0.613 & 1.147 & 0.578 & 0.674 & 0.673 & 0.606 & 0.611$^b$ & 0.620\\
\hline
 & &\multicolumn{14}{c}{$n$-pentane} \\
\hline
TT & C$_{2v}$ & 0.000 & 0.000 & 0.000 & 0.000 & 0.000 & 0.000 & 0.000 & 0.000 & 0.000 & 0.000 & 0.000 & & 0.000 & 0.000\\			
TG & C$_1$    & 1.191 & 0.548 & 0.654 & 0.658 & 0.581 & 0.664 & 0.586 & 1.201 & 0.569 & 0.675 & 0.684 & & 0.614 & 0.613\\ 
GG & C$_2$    & 2.388 & 0.778 & 1.072 & 1.096 & 0.919 & 1.127 & 0.946 & 2.407 & 0.804 & 1.104 & 1.136 & & 0.961 & 0.977\\ 
GX$^-$ & C$_1$   & 4.275 & 2.783 & 2.972 & 2.957 & 2.767 & 3.008 & 2.817 & 4.292 & 2.811 & 3.002 & 3.002 & & 2.813 & 2.833\\
\hline
 & &\multicolumn{14}{c}{$n$-hexane} \\
\hline
TTT & C$_{2h}$ & 0.000 & 0.000 & 0.000 & 0.000 & 0.000  & & & 0.000 & 0.000 & 0.000 & 0.000 &  & 0.000 & 0.000 \\
GTT & C$_1$    & 1.203 & 0.517 & 0.633 & 0.639 & 0.558  & & & 1.213 & 0.539 & 0.656 & 0.666 &  & 0.595 & 0.589 \\
TGT & C$_2$    & 1.249 & 0.517 & 0.642 & 0.649 & 0.561  & & & 1.262 & 0.543 & 0.669 & 0.682 &  & 0.604 & 0.595 \\
TGG & C$_1$    & 2.499 & 0.706 & 1.033 & 1.070 & 0.871  & & & 2.521 & 0.741 & 1.076 & 1.124 &  & 0.934 & 0.930 \\
GTG & C$_2$    & 2.408 & 1.011 & 1.243 & 1.262 & 1.101  & & & 2.428 & 1.057 & 1.292 & 1.320 &  & 1.178 & 1.165 \\
G+T+G- & C$_i$ & 2.481 & 1.179 & 1.399 & 1.398 & 1.240  & & & 2.501 & 1.218 & 1.439 & 1.447 &  & 1.302 & 1.305 \\
GGG & C$_2$    & 3.696 & 0.914 & 1.429 & 1.487 & 1.180  & & & 3.727 & 0.951 & 1.482 & 1.553 &  & 1.250 & 1.260 \\
G+X-T+ & C$_1$ & 4.282 & 2.554 & 2.805 & 2.802 & 2.584  & & & 4.301 & 2.580 & 2.835 & 2.848 &  & 2.632 & 2.646 \\
T+G+X- & C$_1$ & 4.343 & 2.654 & 2.885 & 2.878 & 2.660  & & & 4.363 & 2.699 & 2.934 & 2.942 &  & 2.740 & 2.733 \\
G+X-G- & C$_1$ & 5.492 & 3.143 & 3.498 & 3.504 & 3.209  & & & 5.518 & 3.185 & 3.545 & 3.571 &  & 3.283 & 3.293 \\
X+G-G- & C$_1$ & 5.666 & 2.912 & 3.357 & 3.354 & 3.013  & & & 5.702 & 2.952 & 3.405 & 3.424 &  & 3.083 & 3.105 \\
X+G-X+ & C$_2$ & 7.787 & 4.860 & 5.275 & 5.237 & 4.855  & & & 7.824 & 4.900 & 5.322 & 5.312 &  & 4.925 & 4.947 \\
\hline\hline
\end{tabular}
\begin{flushleft}
$^a$SCF and CCSD energies extrapolated from cc-pV\{T,Q\}Z basis set pair, and the (T) contribution extrapolated from the cc-pV\{D,T\}Z basis set pair.\\
$^b$Using the augmented basis sets results in 0.598 kcal/mol.\\
\end{flushleft}
\end{table}
\clearpage

\squeezetable
\begin{table}
\caption{Comparison between various DFT functionals (pc-2 basis set, without dispersion correction) and our best conformer energies (kcal/mol) for $n$-butane, $n$-pentane, and $n$-hexane.\label{tab:dft-ener}}
\resizebox{1.00\textwidth}{!}{%
\begin{tabular}{l|c|ccccccccccccccc}
\hline\hline
 & Ref.$^a$ & B1B95 & B3LYP & BLYP & M06L & M06 & M06-2X & PBE0 & PBE & PW6B95 & B2GP-PLYP & B2K-PLYP & B2-PLYP & BMK & B3PW91 & MP2$^b$\\
\hline
 & &\multicolumn{15}{c}{$n$-butane} \\
\hline
T & 0.000 & 0.000 & 0.000 & 0.000 & 0.000 & 0.000 & 0.000 & 0.000 & 0.000 & 0.000 & 0.000 & 0.000 & 0.000 & 0.000 & 0.000 & 0.000\\
G & 0.598 & 0.778 & 0.898 & 0.925 & 0.393 & 0.466 & 0.522 & 0.820 & 0.813 & 0.712 & 0.986 & 0.990 & 0.986 & 0.703 & 0.968 & 0.561\\
\hline
 & &\multicolumn{15}{c}{$n$-pentane} \\
\hline
TT & 0.000 & 0.000 & 0.000 & 0.000 & 0.000 & 0.000 & 0.000 & 0.000 & 0.000 & 0.000 & 0.000 & 0.000 & 0.000 & 0.000 & 0.000 & 0.000\\
TG & 0.614 & 0.771 & 0.912 & 0.938 & 0.439 & 0.526 & 0.578 & 0.826 & 0.813 & 0.712 & 0.749 & 0.724 & 0.790 & 0.595 & 0.922 & 0.548\\
GG & 0.961 & 1.239 & 1.665 & 1.737 & 0.339 & 0.449 & 0.581 & 1.444 & 1.433 & 1.079 & 1.383 & 1.294 & 1.537 & 1.048 & 1.665 & 0.778\\
GX$^-$& 2.813 & 3.131 & 3.459 & 3.417 & 2.433 & 2.467 & 2.585 & 3.345 & 3.220 & 2.947 & 3.226 & 3.174 & 3.312 & 2.868 & 3.565 & 2.783\\
\hline
 & &\multicolumn{15}{c}{$n$-hexane} \\
\hline
TTT    & 0.000 & 0.000 & 0.000 & 0.000 & 0.000 & 0.000 & 0.000 & 0.000 & 0.000 & 0.000 & 0.000 & 0.000 & 0.000 & 0.000 & 0.000 & 0.000\\
GTT    & 0.595 & 0.768 & 0.905 & 0.928 & 0.355 & 0.427 & 0.502 & 0.820 & 0.803 & 0.702 & 0.726 & 0.699 & 0.773 & 0.651 & 0.914 & 0.517\\
TGT    & 0.604 & 0.808 & 0.942 & 0.960 & 0.441 & 0.500 & 0.541 & 0.861 & 0.839 & 0.745 & 0.750 & 0.719 & 0.804 & 0.745 & 0.957 & 0.517\\
TGG    & 0.934 & 1.361 & 1.763 & 1.839 & 0.400 & 0.527 & 0.641 & 1.542 & 1.536 & 1.193 & 1.410 & 1.306 & 1.586 & 1.143 & 1.769 & 0.706\\
GTG    & 1.178 & 1.548 & 1.834 & 1.884 & 0.783 & 0.893 & 1.004 & 1.665 & 1.640 & 1.415 & 1.477 & 1.417 & 1.580 & 1.326 & 1.855 & 1.011\\
G+T+G- & 1.302 & 1.608 & 1.871 & 1.908 & 0.845 & 1.065 & 1.178 & 1.714 & 1.669 & 1.479 & 1.563 & 1.517 & 1.641 & 1.454 & 1.890 & 1.179\\
GGG    & 1.250 & 1.823 & 2.595 & 2.728 & 0.390 & 0.594 & 0.751 & 2.238 & 2.252 & 1.561 & 1.983 & 1.822 & 2.258 & 1.586 & 2.597 & 0.914\\
G+X-T+ & 2.632 & 3.019 & 3.421 & 3.390 & 2.241 & 2.303 & 2.339 & 3.260 & 3.148 & 2.820 & 3.066 & 3.001 & 3.177 & 2.919 & 3.502 & 2.554\\
T+G+X- & 2.740 & 3.122 & 3.449 & 3.419 & 2.325 & 2.359 & 2.468 & 3.318 & 3.181 & 2.922 & 3.130 & 3.068 & 3.233 & 2.907 & 3.549 & 2.654\\
G+X-G- & 3.283 & 3.737 & 4.313 & 4.317 & 2.647 & 2.815 & 2.953 & 4.051 & 3.915 & 3.498 & 3.818 & 3.730 & 3.966 & 3.458 & 4.371 & 3.143\\
X+G-G- & 3.083 & 3.713 & 4.376 & 4.422 & 2.216 & 2.432 & 2.631 & 4.072 & 3.947 & 3.386 & 3.725 & 3.613 & 3.915 & 3.270 & 4.497 & 2.912\\
X+G-X+ & 4.925 & 5.566 & 6.227 & 6.207 & 4.199 & 4.480 & 4.514 & 5.957 & 5.734 & 5.199 & 5.593 & 5.501 & 5.750 & 5.179 & 6.406 & 4.860\\
RMSD$^c$&& 0.44 & 0.90 & 0.93 & 0.57 & 0.41 & 0.31 & 0.69 & 0.61 & 0.22 & 0.47 & 0.38 & 0.61 & 0.20 & 0.97 & 0.16\\
\hline
 & &\multicolumn{15}{c}{With standard "-D" dispersion correction} \\
RMSD$^d$&& 0.86 & 0.90 & 1.13 & 0.91 & 0.84 & 0.31 & 0.52 & 0.69 & 0.64 & 0.21 & 0.13 & 0.34 & 0.92 & 0.09 & 0.15\\
$s_6$&& 0.75 & 1.05 & 1.20 & 0.20 & 0.25 & 0.00 & 0.70 & 0.75 & 0.50 & 0.40 & 0.30 & 0.55 & 0.65 & 1.10 & (-0.16)\\
\hline
 & &\multicolumn{15}{c}{With {\em ad hoc} dispersion correction} \\
 RMSD$^e$&& 0.04 & 0.03 & 0.06 & 0.06 & 0.06 & 0.04 & 0.03 & 0.05 & 0.04 & 0.02 & 0.02 & 0.05 & 0.06 & 0.05 & \\
$s_6$&& 0.25 & 0.52 & 0.54 & (-0.33) & (-0.24) & (-0.18) & 0.40 & 0.35 & 0.13 & 0.28 & 0.22 & 0.36 & 0.11 & 0.56 & \\
\hline\hline
\end{tabular}}
\begin{flushleft}
$^a$Best values from Table \ref{tab:ab-initio}: W1h-val throughout.\\
$^b$cc-pVTZ basis set.\\
$^c$Over the $n$-hexane conformers, without dispersion correction.\\
$^d$Over the $n$-hexane conformers, with dispersion correction using standard s$_6$ values for the functionals (see text) given in the subsequent row.\\
$^e$Over the $n$-hexane conformers, with dispersion correction using {\em ad hoc} optimized s$_6$ values given in the subsequent row. Negative $s_6$ 
values can be seen as "undoing" an overcorrection in the underlying level of theory.\\
\end{flushleft}
\end{table}
\clearpage

\squeezetable
\begin{table}
\caption{Backbone torsion angles (degree) of the $n$-pentane conformers obtained with different DFT functionals in conjunction with the pc-2 basis set.\label{tab:dihedral}}
\begin{tabular}{l|l|cccccccccccc}
\hline\hline
Conformer & $\tau$  & B1B95 & B3LYP & B3PW91 & BLYP & BMK & M06-2X & M06-L & M06 & PBE0 & PBE & PW6B95 & MP2$^a$ \\
\hline
TG  & $\tau_1$  & 177.1 & 177.3 & 177.3 & 177.4 & 177.0 & 175.3 & 177.2 & 175.2 & 177.2 & 177.3 & 176.9 & 176.2 \\ 
    & $\tau_2$  & 64.7 & 66.2 & 66.0 & 66.8 & 64.1 & 61.6 & 62.3 & 62.7 & 65.3 & 65.7 & 64.5 & 64.0 \\
GG  & $\tau_1$  & 56.5 & 63.6 & 63.1 & 64.9 & 58.5 & 56.6 & 58.0 & 58.5 & 61.7 & 62.8 & 56.5 & 58.3 \\
GX$^-$ & $\tau_1$  & -98.3 & -90.9 & -90.9 & -89.3 & -97.8 & -95.8 & -92.5 & -94.4 & -93.9 & -91.7 & -98.0 & -96.4 \\
    & $\tau_2$  & 59.3 & 65.0 & 65.0 & 67.1 & 59.1 & 58.6 & 59.2 & 61.8 & 62.3 & 64.1 & 59.3 & 59.9 \\
\hline\hline
\end{tabular}
\begin{flushleft}
$^a$cc-pVTZ basis set.\\
\end{flushleft}
\end{table}

\squeezetable
\begin{table}
\caption{Relative energies, point groups, and degeneracies of the $n$-octane, $n$-heptane, isohexane, 3-methylpentane, isooctane, isoheptane, neopentane, and isopentane conformers at the B2K-PLYP-D(0.22)/pc-2//PW6B95/6-311G(d,p) level of theory (in kcal/mol).\label{tab:b2k-plyp}}
\resizebox{1.00\textwidth}{!}{%
\begin{tabular}{clcc|clcc|clcc|clcc}
\hline\hline
Degen. & Conformer & Symmetry & Energy & Degen. & Conformer & Symmetry & Energy & Degen. & Conformer & Symmetry & Energy & Degen. & Conformer & Symmetry & Energy\\
\hline
\multicolumn{4}{c}{\bf n-octane} & \multicolumn{4}{c}{\bf n-octane} & \multicolumn{4}{c}{\bf n-heptane} & \multicolumn{4}{c}{\bf isoheptane}\\
 1& TTTTT     &C$_{2h}$& 0.000 & 4 & G-G-TG-X+ & C$_1$ & 3.572  & 1 & TTTT    &C$_{2v}$& 0.000  & 2 & TTG-  & C$_1$ & 0.000\\
 4& TTTTG-    & C$_1$  & 0.584 & 4 & G+TG+G+X- & C$_1$ & 3.416  & 4 & TTTG-   & C$_1$  & 0.581  & 2 & TG+T  & C$_1$ & 0.306\\
 2& TTG-TT    & C$_2$  & 0.545 & 4 & G-G-TX-G+ & C$_1$ & 3.538  & 4 & TTG-T   & C$_1$  & 0.570  & 1 & TTG+  & C$_s$ & 0.778\\
 4& TTTG-T    & C$_1$  & 0.556 & 4 & TTG-X+G+  & C$_1$ & 3.164  & 4 & TTG-G-  & C$_1$  & 0.916  & 2 & G+TT  & C$_1$ & 0.539\\
 4& TTG-G-T   & C$_1$  & 0.822 & 4 & TX+G-TG+  & C$_1$ & 3.274  & 2 & TG+G+T  & C$_2$  & 0.860  & 2 & G+G+T & C$_1$ & 0.600\\
 4& TTTG-G-   & C$_1$  & 0.890 & 4 & TG+G+G+X- & C$_1$ & 3.369  & 4 & TG+TG+  & C$_1$  & 1.140  & 2 & G+TG- & C$_1$ & 0.690\\
 4& TTG-TG-   & C$_1$  & 1.098 & 4 & TTG-X-G+  & C$_1$ & 3.262  & 2 & G+TTG+  & C$_2$  & 1.148  & 2 & G+TG+ & C$_1$ & 1.445\\
 2& TG+TG+T   & C$_2$  & 1.099 & 4 & TG+X-G-T  & C$_1$ & 3.136  & 2 & G+TTG-  & C$_s$  & 1.152  & 2 & TX+G- & C$_1$ & 2.047\\
 2& G+TTTG+   & C$_2$  & 1.159 & 4 & G+G+TX-G+ & C$_1$ & 3.594  & 4 & TG+TG-  & C$_1$  & 1.265  & 2 & TG+X- & C$_1$ & 2.426\\
 4& TG+TTG+   & C$_1$  & 1.118 & 4 & TX+G-G-G- & C$_1$ & 3.224  & 4 & TG+G+G+ & C$_1$  & 1.186  & 2 & TG+G+ & C$_1$ & 3.025\\
 2& G+TTTG-   & C$_i$  & 1.156 & 4 & TG+TG-X+  & C$_1$ & 3.424  & 4 & G+TG+G+ & C$_1$  & 1.447  & 2 & X+G-G-& C$_1$ & 2.480\\
 4& TTG-G-G-  & C$_1$  & 1.123 & 4 & G+G+X-TG- & C$_1$ & 3.431  & 2 & G+TG-G- & C$_1$  & 1.606  & 3 & TX+G+ & C$_1$ & 2.784\\
 2& TG+G+G+T  & C$_2$  & 1.071 & 4 & G+TG-G-X+ & C$_1$ & 3.567  & 4 & G+G+G+G+& C$_2$  & 1.519  & 2 & G+X-G-& C$_1$ & 2.567\\
 4& TG+TTG-   & C$_1$  & 1.117 & 4 & G+G+TG-X+ & C$_1$ & 3.724  & 4 & TTX-G+  & C$_1$  & 2.674  & 2 & G+G+C+& C$_1$ & 3.404\\
 4& TG+G+TG+  & C$_1$  & 1.343 & 4 & TG+G+X-G- & C$_1$ & 3.192  & 4 & TTG-X+  & C$_1$  & 2.740  & 2 & G+X+G-& C$_1$ & 2.627\\
 4& TG+TG+G+  & C$_1$  & 1.368 & 4 & G+TX-G+G+ & C$_1$ & 3.591  & 4 & TG+X-T  & C$_1$  & 2.608  & 2 & G+G+X-& C$_1$ & 2.779\\
 4& TTG-TG+   & C$_1$  & 1.229 & 4 & G+G+G+G+X-& C$_1$ & 3.702  & 4 & TG+G+X- & C$_1$  & 2.956  & 2 & G+X+G+& C$_1$ & 3.466\\
 4& G+TTG+G+  & C$_1$  & 1.430 & 4 & G+TG+X-G- & C$_1$ & 3.778  & 4 & TX+G-G- & C$_1$  & 2.917  & 2 & X+G-T-& C$_1$ & 4.577\\
 4& G-TTG+G+  & C$_1$  & 1.434 & 4 & TG+X+G-G- & C$_1$ & 3.279  & 4 & G+TX+G- & C$_1$  & 3.261  & \multicolumn{4}{c}{\bf neoheptane} \\
 2& TG+TG-T   & C$_i$  & 1.246 & 4 & G+TG+X+G- & C$_1$ & 3.795  & 8 & G+TG+X- & C$_1$  & 3.329  & 1 & T     & C$_s$ & 0.000\\
 2& G+TG+TG+  & C$_2$  & 1.685 & 4 & G+TX-X-G+ & C$_1$ & 3.821  & 4 & G+TX-G+ & C$_1$  & 3.332  & 2 & X-    & C$_1$ & 2.256\\
 2& G+G+TG+G+ & C$_2$  & 1.656 & 4 & TG+X-G-G- & C$_1$ & 3.793  & 8 & TG+X+G- & C$_1$  & 3.280  & 2 & G-    & C$_1$ & 2.708\\
 4& TG+G+G+G+ & C$_1$  & 1.413 & 4 & G-X-G+G+G+& C$_1$ & 3.736  & 4 & TG+X-G- & C$_1$  & 3.220  & \multicolumn{4}{c}{\bf isopentane}\\
 4& G-G-G-TG- & C$_1$  & 1.634 & 4 & G+TG-X+G+ & C$_1$ & 3.881  & 4 & G+G+G+X-& C$_1$  & 3.506  & 2 & G+   & C$_1$ & 0.000\\
 4& TG+G+TG-  & C$_1$  & 1.495 & 4 & G+G+G+X+G-& C$_1$ & 4.115  & 4 & G+G+X-G-& C$_1$  & 3.413  & 1 & G-   & C$_s$ & 0.786\\
 4& TG+TG-G-  & C$_1$  & 1.574 & 4 & G+G+X-G-G-& C$_1$ & 3.863  & 2 & TX+G-X+ & C$_1$  & 4.636  & \multicolumn{4}{c}{\bf isooctane}\\
 4& G+TG-TG-  & C$_1$  & 1.822 & 4 & TG+X-X-G+ & C$_1$ & 4.804  & 4 & G+X-X-G+& C$_2$  & 4.897  & 2 & X-    & C$_1$ & 0.000\\
 2& G+G+G+G+G+& C$_2$  & 1.752 & 2 & TX+G-X+T  & C$_2$ & 4.223  & 4 & G+X+G-X+& C$_1$  & 5.426  & 2 & G-    & C$_1$ & 0.483\\
 4& G+TG-G-G- & C$_1$  & 1.805 & 4 & TTX-G+X-  & C$_1$ & 4.560  & 4 & L+G-X-G+& C$_1$  & 6.051  & 2 & X+    & C$_1$ & 3.319\\
 2& G+G+TG-G- & C$_i$  & 1.908 & 2 & G+X-TX-G+ & C$_2$ & 5.212  & 2 & X+G-G-X+& C$_2$  & 6.288  &  &       &       & \\
 2& G+TG-TG+  & C$_2$  & 1.959 & 4 & G+G+X-X-G+& C$_1$ & 5.183  &  \multicolumn{4}{c}{\bf isohexane}&&    &       & \\
 4& TTTX-G+   & C$_1$  & 2.658 & 2 & X+G-TG-X+ & C$_2$ & 5.508  & 2 & TG-     & C$_1$  & 0.000 &  &       &       & \\
 4& TTG-X+T   & C$_1$  & 2.507 & 4 & G+TX+G-X+ & C$_1$ & 5.248  & 2 & G+T     & C$_1$  & 0.352 &  &       &       & \\
 4& TTX-G+T   & C$_1$  & 2.561 & 4 & X+G-TX+G- & C$_1$ & 5.469  & 1 & TG+     & C$_s$  & 0.803 &  &       &       & \\
 4& TTTG-X+   & C$_1$  & 2.713 & 4 & G+TX-G+X- & C$_1$ & 5.255  & 2 & X+G-    & C$_1$  & 2.220 &  &       &       & \\
 4& TTG-G-X+  & C$_1$  & 2.883 & 4 & X+G-TX-G+ & C$_1$ & 5.488  & 2 & G+X-    & C$_1$  & 2.533 &  &       &       & \\
 4& TX+G-TG-  & C$_1$  & 3.140 & 4 & TX+G-X+G+ & C$_1$ & 5.078  & 2 & G+G+    & C$_1$  & 3.165 &  &       &       & \\
 4& TX+G-G-T  & C$_1$  & 2.702 & 2 & G+X-TX+G- & C$_i$ & 5.544  & 2 & X+G+    & C$_1$  & 2.983 &  &       &       & \\
 4& TG+TG+X-  & C$_1$  & 3.303 & 4 & TX+X+G-X+ & C$_1$ & 5.328  & \multicolumn{4}{c}{\bf 3-methylpentane}&&&     & \\
 4& G+TTG+X-  & C$_1$  & 3.255 & 2 & X+G-G-G-X+& C$_2$ & 5.544  & 2 & TG      & C$_1$  & 0.353 &  &       &       & \\
 4& G+TTX-G+  & C$_1$  & 3.171 & 2 & X+G-TG+X- & C$_i$ & 5.700  & 2 & G-T     & C$_1$  & 0.115 &  &       &       & \\
 4& TTX-G+G+  & C$_1$  & 2.880 & 4 & X+G-G-X-G+& C$_1$ & 5.803  & 2 & G-G-    & C$_1$  & 0.630 &  &       &       & \\
 4& G+TTX+G-  & C$_1$  & 3.228 & 2 & G+X-X-X-G+& C$_2$ & 5.914  & 1 & G-G+    & C$_s$  & 0.000 &  &       &       & \\
 4& TG+X-TG-  & C$_1$  & 3.127 & 4 & TG+X-G-L+ & C$_1$ & 5.913  & 2 & T-T     & C$_1$  & 1.862 &  &       &       & \\
 4& G+TTG-X+  & C$_1$  & 3.254 & 4 & G+G+X+G-X+& C$_1$ & 6.128  & 2 & C+G-    & C$_1$  & 3.222 &  &       &       & \\
 4& TG+TX+G-  & C$_1$  & 3.213 & 2 & G+X+G-X+G+& C$_2$ & 5.942  &  &         &        &       &  &       &       & \\
 4& TG+X-TG+  & C$_1$  & 3.237 & 4 & X+G-X+X+G-& C$_1$ & 6.991  &  &         &        &       &  &       &       & \\
 4& TG+TX-G+  & C$_1$  & 3.287 & 4 & L+G-X-G+X-& C$_1$ & 8.083  &  &         &        &       &  &       &       & \\
\hline\hline
\end{tabular}}
\begin{flushleft}
\end{flushleft}
\end{table}
\clearpage

\squeezetable
\begin{table}
\caption{Enthalpy function (H$_{298}-$H$_0$) and Gibbs energy function calculated with various DFT functionals with the pc-2 basis set.\label{tab:ef-gef}}
\begin{tabular}{l|ccccccccccccc}
\hline\hline
& B1B95 & B2K-PLYP & B2K-PLYP-D & B3LYP & BLYP & M06-2X & M06 & M06-L & PBE & PBE0 & PW6B95 & CCSD(T) & W1h-val$^a$ \\
\hline
& \multicolumn{12}{c}{H$_{298}-$H$_0$  (kcal/mol)}\\
\hline
n-butane          & 0.272 & 0.269 & 0.256 & 0.274 & 0.273 & 0.237 & 0.222 & 0.199 & 0.274 & 0.274 & 0.267 & 0.254 & 0.252\\
n-pentane         & 0.513 & 0.497 & 0.468 & 0.514 & 0.514 & 0.435 & 0.395 & 0.336 & 0.514 & 0.513 & 0.504 & 0.462 & 0.473\\
isopentane        & 0.090 & 0.090 & 0.092 & 0.085 & 0.083 & 0.093 & 0.093 & 0.087 & 0.086      & 0.087      & 0.091  & \\
n-hexane          & 0.757 & 0.726 & 0.678 & 0.748 & 0.746 & 0.608 & 0.546 & 0.464 & 0.750 & 0.750 & 0.746 & 0.663 & 0.686\\
isohexane         & 0.267 & 0.269 & 0.243 &       &       & 0.164 & 0.170 &0.148  & 0.278      &  0.271     &  0.261 & \\
3-methylpentane   & 0.308 & 0.303 & 0.265 & 0.326 & 0.322 & 0.168 & 0.150 & 0.154 & 0.327 & 0.326 & 0.294 & \\
diisopropyl       & 0.009 & 0.055 & 0.041 &    0.025   & 0.018 & 0.037 &-0.075 & -0.164 & 0.055      & 0.056      & 0.003  & \\
n-heptane         & 1.005 & 0.960 & 0.886 & 0.942 & 0.921 & 0.840 & 0.837 & 0.649 & 0.966 & 0.974 & 0.991 & & 0.893$^b$\\
isoheptane        & 0.520 & 0.513 & 0.456 & 0.583 & 0.487 & 0.353 & 0.401 & 0.286 & 0.530      & 0.525      & 0.506  & \\
neoheptane        & 0.160 & 0.129 & 0.147 &       & 0.138 & 0.223 & 0.221 & 0.257 &       & 0.115      & 0.183  & \\
n-octane          & 1.250 & 1.190 & 1.076 & 1.178 & 1.153 & 0.946 & 0.942 & 1.250 & 1.209 & 1.250 & 1.222 & \\
isooctane         & 0.167 & 0.161 & 0.156 &0.169 &0.170 &0.147 &0.128 &0.134 &0.167 &0.167&0.164 & \\
\hline
& \multicolumn{12}{c}{Gibbs energy function (cal/K.mol)}\\
\hline
n-butane          & 0.856 & 0.910 & 1.053 & 0.724 & 0.696 & 1.199 & 1.286 & 1.407 & 0.815 & 0.808 & 0.935 & 1.073 & 1.088\\
n-pentane 		   & 1.704 & 1.758 & 2.090 & 1.367 & 1.317 & 2.379 & 2.570 & 2.805 & 1.576 & 1.550 & 1.864 & 2.135 & 2.080\\
isopentane 		   & 1.592 & 1.596 & 1.625 & 1.556 & 1.546 & 1.651 & 1.684 & 1.778 & 1.560      &  1.569     & 1.606  & \\
n-hexane 		   & 2.454 & 2.632 & 3.166 & 1.982 & 1.917 & 3.654 & 3.927 & 4.264 & 2.283 & 2.238 & 2.708 & 3.286 & 3.148\\
isohexane 		   & 2.265 & 2.215 & 2.466 &       &       & 2.851 & 2.849 & 2.997 & 2.047      &  2.066     & 2.378  & \\
3-methylpentane    & 2.808 & 2.797 & 3.004 & 2.543 & 2.506 & 3.408 & 3.471 & 3.458 & 2.579 & 2.614 & 2.897 & \\
diisopropyl  	   & 2.153 & 1.993 & 2.043 &   2.099    &  2.124     & 2.056 & 2.427 & 2.706 & 1.993      & 1.991      & 2.172  & \\
n-heptane 		   & 3.262 & 3.512 & 4.284 & 2.392 & 2.229 & 4.693 & 4.795 & 5.499 & 2.850 & 2.834 & 3.603 & & 4.256$^b$\\
isoheptane 	   	   & 3.093 & 3.118 & 3.574 & 2.452 & 2.343 & 4.008 & 3.957 & 4.410 & 2.693      & 2.724      & 3.292  & \\
neoheptane 		   & 0.145 & 0.105 & 0.125 &       & 0.113 & 0.232 & 0.225 & 0.288 &       & 0.091      & 0.176  & \\
n-octane 		   & 4.091 & 4.436 & 5.451 & 2.946 & 2.720 & 6.186 & 6.346 & 4.091 & 3.552 & 4.091 & 4.559 & \\
isooctane 		   & 1.994 & 2.055 & 2.111 &1.943& 1.910 &2.196 &2.313 &2.313&1.980&1.981&2.038 & \\
\hline\hline
\end{tabular}
\begin{flushleft}
B2K-PLYP and B2GP-PLYP results at PW6B95/6-311G(d,p) geometries. Geometries fully optimized at remaining levels of theory for butane, pentane, hexane, and neoheptane, but PW6B95/6-311G(d,p) reference geometries used for remaining heptanes and octanes.\\
CCSD(T) results at MP2/cc-pVTZ geometries, using AVQZ basis set for n-butane, AVTZ basis set for n-pentane, and PVTZ basis set for n-hexane.\\
$^a$SCF and CCSD energies extrapolated from cc-pV\{T,Q\}Z basis set pair, and the (T) contribution extrapolated from the cc-pV\{D,T\}Z basis set pair.
$^b$MP2:CC result.
\end{flushleft}
\end{table}
\clearpage

\squeezetable
\begin{table}
\caption{B2K-PLYP-D(0.22)/pc-2//PW6B95/6-311G(d,p) thermal conformer corrections at 298.15K using three approximations: (a) using  `bottom of the well' conformer energy differences; (b) using the same at 0 K; (c) accounting for individual rovibrational partition functions.\label{tab:thermal-corrections}}
\begin{tabular}{l|ccc|ccc}
\hline\hline
 & \multicolumn{3}{c}{gef(T) in e.u.} & \multicolumn{3}{c}{H$_{298}-$H$_0$ in kcal/mol}\\
 & just $\Delta E_e$ & just $\Delta E_0$ & vib.avg. & just $\Delta E_e$  & just $\Delta E_0$ & vib.avg.\\
\hline 
$n$-butane & 1.053 & 1.067 & 1.149 & 0.256 & 0.255 & 0.267\\
$n$-pentane & 2.090 & 1.748 & 2.539 & 0.468 & 0.489 & 0.508\\
$n$-hexane & 3.166 & 2.651 & 3.419 & 0.678 & 0.718 & 0.682\\
$n$-heptane & 4.284 & 3.632 & 4.628 & 0.886 & 0.934 & 0.826\\
$n$-octane & 5.466& 4.528& 5.301& 1.081 & 1.193 &1.025\\
3-methylpentane & 3.004 & 2.632 & 2.429 & 0.265 & 0.324 & 0.229\\
diisopropyl & 2.043 & 2.088 & 1.981 & 0.041 & 0.028 & 0.031\\
isoheptane & 3.573 & 3.086 & 2.761 & 0.456 & 0.516 & 0.401\\
isohexane & 2.466 & 2.149 & 1.893 & 0.243 & 0.271 & 0.173\\
isooctane & 2.111 & 2.000 & 1.849 & 0.156 & 0.165 & 0.122\\
isopentane & 1.625 & 1.568 & 1.568 & 0.092 & 0.087 & 0.084\\
neoheptane & 0.125 & 0.075 & 0.027 & 0.147 & 0.100 & 0.033\\
\hline
\multicolumn{7}{c}{Linear regression for n-alkanes (\# of backbone torsions)}\\
\hline
slope & 1.102 & 0.881 & 1.039 & 0.207 & 0.232 & 0.183\\
intercept & -0.094 & 0.083 & 0.289 & 0.053 & 0.021 & 0.111\\
$R^2$ & 0.9993 & 0.9964 & 0.9882 & 0.9998 & 0.9993 & 0.9928\\
\hline
\multicolumn{7}{c}{With individual internal rotation corrections for each conformer}\\
\hline
$n$-butane  & 1.030 & 1.043 & 1.125 & 0.259 & 0.257 & 0.270\\
$n$-pentane & 2.138 & 1.778 & 2.557 & 0.466 & 0.493 & 0.508\\
$n$-hexane  & 3.246 & 2.704 & 3.458 & 0.672 & 0.723 & 0.682\\
$n$-heptane & 4.408 & 3.716 & 4.686 & 0.873 & 0.939 & 0.824\\
\hline\hline
\end{tabular}
\begin{flushleft}
\end{flushleft}
\end{table}

\end{document}